\documentclass[prc,superscriptaddress,twocolumn,showpacs,preprintnumbers,amsmath,amssymb]{revtex4-1}
\usepackage{amsmath}
\usepackage{times}
\usepackage{braket}

\def\fun#1#2{\lower3.6pt\vbox{\baselineskip0pt\lineskip.9pt
 \ialign{$\mathsurround=0pt#1\hfil##\hfil$\crcr#2\crcr\sim\crcr}}}
\usepackage{graphicx}
\usepackage{dcolumn}
\usepackage{bm}

\usepackage{ulem}
\usepackage{color}
\definecolor{verde}{rgb}{0.0,0.7,0.0}

\begin{document}


\title{Spin-Isospin Properties of $\boldsymbol{N=Z}$ Odd-Odd Nuclei from a Core+$\boldsymbol{pn}$ Three-Body Model including Core Excitations}

\author{Futoshi Minato}
\email{minato.futoshi@jaea.go.jp}
\affiliation{Nuclear Data Center, Japan Atomic Energy Agency, Tokai, Ibaraki 319-1195, Japan}
\affiliation{NSCL/FRIB Laboratory, Michigan State University, East Lansing, Michigan 48824, USA}

\author{Yusuke Tanimura}
\email{tanimura@nucl.phys.tohoku.ac.jp}
\affiliation{Department of Physics, Tohoku University, Sendai, 980-8578, Japan}

\date{\today}

\begin{abstract}
 \noindent
 \textbf{Background:}
For $N=Z$ odd-odd nuclei, a three-body model assuming two valence particles and an inert core can provide an understanding of pairing correlations in the ground state and spin-isospin excitations. However, since residual core-nucleon interactions can have a significant impact on these quantities, the inclusion of core excitations in the model is essential for useful calculation to be performed.
\\
 \textbf{Purpose:}
The effect of core excitations must be included in order to gain a detailed understanding of both the ground state and spin-isospin properties of these systems. To this end, we include the vibrational excitation of the core nucleus in our model. 
\\
 \textbf{Methods:}
We solve the three-body core-nucleon-nucleon problem including core vibrational states to obtain the nuclear ground state as well as spin-isospin excitations. The core vibrational states are described by the random-phase-approximation. The spin-isospin excitations are examined from the point of view of SU(4) multiplets.
\\
 \textbf{Results:}
By including the effect of core excitation, the following experimental quantities of $N=Z$ odd-odd nuclei are better described: (a) the magnetic moment of the ground states, (b) the energy difference between the first $0^+$ and $1^+$ states, and (c) the $B$($M1$) and $B$(GT) between $0^+$ and $1^+$ states. In addition, the coupling with the core vibration induces quenching and damping in $B$($M1$) and $B$(GT), and the root mean square distances between proton and neutron and that between the center of mass of proton and neutron and core nucleus increase. Large $B$($M1$) and $B$(GT) observed for $^{18}$F and $^{40}$Ca were explained in terms of the SU(4) symmetry.
 \\
 \textbf{Conclusions:}
The core nucleus is meaningfully broken by the residual core-nucleon interactions, and various quantities concerning spin-isospin excitations as well as the ground state become consistent with experimental data. Including the core excitation in the three-body model is thus important for a more detailed understanding of nuclear structure.
\end{abstract}

\pacs{}

\maketitle

\section{Introduction}
\label{Intro}
%
The nuclear many-body problem remains a complicated and difficult task to solve due to the large number of degrees of freedom involved.
However, schematic models simplifying the nuclear many-body problem have succeeded in explaining experimental observables theoretically. 
The typical method is a three-body model, which assumes two particles and a core nucleus (or three particles) in the system. The core is usually either a closed-shell nucleus or one which can be virtually regarded as structureless. This assumption greatly reduces the model space greatly and allows us to solve the many-body problem rather simply. Thus, the three-body model has been applied to various nuclear systems, for example, $^{6}$Li (proton, neutron, and $^{4}$He), weakly bound systems such as $^{6}$He (two neutrons and $^{4}$He) and $^{11}$Li (two neutrons and $^{9}$Li)~\cite{Bertsch1991, Zhukov1991, Esbensen1992}, one-proton radioactivity of $^6_\Lambda$Li hypernucleus~\cite{Oishi2018} and deuteron clusterization~\cite{SMWang, Oishi2017}. 
In the last decade, new experimental facilities using a highly intensive radioactive beam have enabled us to explore nuclei near the drip-line of the nuclear chart, where the three-body model becomes an effective approach to study di-neutron and di-proton correlations in exotic nuclei~\cite{Hagino2005, Hagino2007, Oishi2010,erratumOishi2010} and quasi-bound states of $^{26}$O~\cite{SMWang, Hagino2014, Hagino2014b, Hagino2016}.

The three-body model also has an advantage to focus on the two-body subsystem consisting of two valence particles and provides us a beneficial understanding of pairing correlations in the ground state and spin-isospin excitations. For this reason, in the previous work~\cite{Tanimura2014, Tanimura2016}, $N=Z$ odd-odd nuclei, which are made up of one proton, one neutron, and an even-even core nucleus with mass $A=14$ to $100$, were investigated in order to study the effect of isospin $T=0$ and $T=1$ pairing correlations between the proton and neutron. It was found that the SU(4) multiplets \cite{Wigner1937}, which are characterized by the nucleon spin-isospin plane, are remarkably present in $^{18}$F and $^{42}$Sc and result in strong magnetic dipole (M1) and Gamow-Teller (GT) transitions.

The assumption of an inert core in the three-body model~\cite{Tanimura2014, Tanimura2016} looks reasonable as long as the energy region in question is below the first excited state of the core nucleus. However, it has been seen that a residual two-body interaction significantly breaks the stationary structure of a core nucleus even for spherical magic nuclei~\cite{Honma2004, Utsuno2011}. Thus, the assumption of the inert core might be too simple. In fact, some physical quantities were not reproduced well in the previous work~\cite{Tanimura2014, Tanimura2016}. One is able to consider the effect of core breaking by extending the active model space to include one or more lower shells and performing a large scale shell model calculation. However, such an approach sacrifices the simple picture of three-body model. As an alternative approach, one can assume the core breaking as a core collective excitation. This approach keeps the concept of a three-body model and allows us to focus on the subsystem consisting of valence nucleons in the same manner as the three-body model with an inert core.

The core collective excitations are characterized as rotation and vibration. The coupling to rotationally excited states have been investigated in the two-body system, i.e. core nucleus + one valence particle~\cite{Nunes1996b, Ridikas1998, Esbensen1995, Esbensen2000, Urata2011} and the three-body system for $^{12}$Be~\cite{Nunes1996, Nunes2002} and $^{11}$Li~\cite{Thompson1994}. The coupling to surface vibrations can be similarly described by particle-vibration coupling~\cite{BohrMottelson, Mahaux1985} and has been applied to various systems, such as light nuclei~\cite{Mau1995}, neighbors magic nuclei~\cite{Hamamoto1969, Hamamoto1973, Hamamoto1976, Bernard1980, SagawaBrown1984, Litvinova2006, Colo2010, Litvinova2011, Cao2014}, open shell nuclei~\cite{Soloviev1987}, and deformed systems~\cite{Yoshida2009}. The valence particle(s) may couple with a pairing vibration, however, its contribution is considered to be small~\cite{Bertsch1979}. While much work has been done studying particle-vibration coupling in even-even nuclei and two-body systems as mentioned above, it has not been studied yet in the three-body system.

We discuss in this paper spin-isospin properties of $N=Z$ nuclei from the viewpoint of one step beyond Refs.~\cite{Tanimura2014, Tanimura2016} by allowing the core nucleus to vibrationally excite. The same nuclei of Ref.~\cite{Tanimura2014} are chosen as the target of this work, that is $^{14}$N ($^{12}$C+$pn$), $^{18}$F ($^{16}$O+$pn$), $^{30}$P ($^{28}$Si+$pn$), $^{34}$Cl ($^{32}$S+$pn$), $^{42}$Sc ($^{40}$Ca+$pn$), and $^{58}$Cu ($^{56}$Ni+$pn$) $N=Z$ odd-odd nuclei. The core vibrational degree of freedom is included in the same manner as used in the particle-vibration coupling of the two-body system~\cite{BohrMottelson, SagawaBrown1984, Hamamoto1996, Yoshida2009}. The obtained result is compared with that calculated by the three-body model using an inert core. Note that $^{14}$N and $^{30}$P have the deformed cores of $^{12}$C and $^{28}$Si, respectively. In these nuclei, rotational coupling will also be present, and so our results are thus a minimum estimation of the coupling with core excitations for those nuclei.

The inclusion of the core vibration will generate some changes in the nuclear structure, which has not been seen in the three-body model with an inert core. In addition to the spin-isospin properties, we therefore discuss mean values of spatial distribution of valence particles. We also clarify which phonon state of the core nucleus plays an important role in the ground state wave function.

Section \ref{theory} describes the framework of three-body model calculation with core excitation and the model space used, and sec.~\ref{result} provides results of magnetic moments, spin-isospin excitations, low-lying GT transitions, core contribution, and mean values of spatial distributions calculated in order to see the effect of core vibrations. The results are compared with those calculated by the three-body model using an inert core. We summarize our discussion in sec.~\ref{summary}.

\section{Theoretical Framework}
\label{theory}

\subsection{Three-body system with core vibrational excitations}
\label{theory1}

The formalism of the three-body model with an inert core can be found in Refs.~\cite{Bertsch1991, Esbensen1997, Hagino2005, Hagino2007}. We now extend it to the case with core vibrational excitations. For general purpose, we label one of the valence nucleons as $1$ and another as $2$ ($1$ and $2$ are either proton or neutron). In the rest frame of the core nucleus, the Hamiltonian is given by
\begin{equation}
\begin{split}
H&=\frac{\vec{p}_1^2}{2m_1}+\frac{\vec{p}_2^2}{2m_2}+V_{1c}(\vec{r}_1)+V_{2c}(\vec{r}_2)
+V_{p}(\vec{r}_1,\vec{r}_2)\\
&+\frac{(\vec{p}_1+\vec{p}_2)^2}{2M_{core}}+H_{core}.
\end{split}
\label{3bodyH}
\end{equation}
The first two terms in Eq.~\eqref{3bodyH} are the kinetic energies, and $m_1$ and $m_2$ take either neutron mass, $m_n$, or proton mass, $m_p$. The function $V_{ic}(\vec{r}_i)$ is a core-nucleon effective interaction working between the valence nucleon labeled by $i$ and the core nucleus, and $V_{p}(\vec{r}_1,\vec{r}_2)$ is a pairing interaction between the valence nucleons (see Eq.~\eqref{NNforce}). The sixth term is the recoil kinetic energy term and $M_{core}=N_c m_n+Z_c m_p$ is the mass of the core nucleus, where $N_c$ and $Z_c$ are the number of neutron and proton of the core nucleus. The last term in Eq.~\eqref{3bodyH} is the Hamiltonian for the intrinsic degree of freedom of the core nucleus.

The effective interaction $V_{ic}(\vec{r}_i)$ is generally density-dependent. Since the amplitude of the core surface vibration is small, we can expand the interaction up to the first-order with respect to the density,
\begin{equation}
V_{ic}(\vec{r})\simeq V_{ic}^0(\vec{r})+\sum_\nu \int \frac{\delta{V_{ic}(\vec{r})}}{\delta\rho(\vec{r}')} 
\delta\hat{\rho}_\nu(\vec{r}')d\vec{r}',
\label{funcderiv}
\end{equation}
The first term is the core-nucleon interaction at the static density of core nucleus, $\delta{V_{ic}(\vec{r})}/\delta\rho(\vec{r}')$ in the second term is the residual interaction generated by the core surface vibration, and $\delta\hat{\rho}_\nu(\vec{r})$ is the transition density operator of the vibrational phonon state $\nu$, acting on a core intrinsic wave function. The tensor operators altering the orbital angular momentum of valence nucleons and core nucleus are included in the transition density operator. The residual interaction can be classified by isoscalar (IS) and isovector (IV) components~\cite{SagawaBrown1984,Hamamoto1996} as
\begin{equation}
\int \frac{\delta V_{ic}(\vec{r})}{\delta\rho(\vec{r}')}\delta\hat{\rho}_\nu(\vec{r}')d\vec{r}' 
=\int \Big(
v_{res}^{\rm{IS}}(\vec{r},\vec{r}')\delta\hat{\rho}_\nu^{\rm{IS}}
+v_{res}^{\rm{IV}}(\vec{r},\vec{r}')\delta\hat{\rho}_\nu^{\rm{IV}}
\Big) d\vec{r}'.
\label{residualISIV}
\end{equation}
The first term of the right hand side of Eq.~\eqref{residualISIV} does not change the isospin of either the $pn$-subsystem or core nucleus, while the second term of the isovector component may change their isospin by $\Delta T=1$. Due to this, for example, the $1^+$ ground state has a component of isospin $T=1$ in both the $pn$-subsystem and core nucleus in addition to the $T=0$ component.

We neglect the recoil kinetic energy of the core nucleus in the sixth term of Eq.~\eqref{3bodyH} as done in Ref.~ \cite{Tanimura2014}. From Eqs.~\eqref{3bodyH}, ~\eqref{funcderiv},~\eqref{residualISIV}, the effective Hamiltonian then becomes
\begin{equation}
\begin{split}
H&= H_1+H_2+V_{p}(\vec{r}_1,\vec{r}_2)+H_{core}\\
&+\sum_{\nu,\beta} \int \left(v_{res}^{\beta}(\vec{r}_1,\vec{r})+v_{res}^{\beta}(\vec{r}_2,\vec{r})\right)\delta\hat{\rho}_\nu^\beta(\vec{r})d\vec{r},
\end{split}
\label{3bodyH2}
\end{equation}
($\beta=$IS and IV), and we define
\begin{equation}
H_i=\frac{\vec{p}_i^2}{2m_i}+V_{ic}^0(\vec{r_i}).
\label{subH}
\end{equation}
Eq.~\eqref{subH} is identical to the single particle shell model Hamiltonian for the valence nucleon $i(=1,2)$ in the static core. By imposing spherical symmetry on the system, it satisfies the relation; 
\begin{equation}
H_i\psi_{njl}(\vec{r}_i)=\varepsilon_{njl}\psi_{njl}(\vec{r}_i),
\label{sph.spl}
\end{equation}
where $\psi_{njl}$ and $\varepsilon_{njl}$ are the eigenfunction and the eigenvalue of the single particle states (principal quantum number $n$, total angular momentum $j$, and orbital angular momentum $l$), respectively.

We diagonalize the effective Hamiltonian of Eq.~\eqref{3bodyH2} by the superposed wave functions of two-body and intrinsic core bases coupled with total spin $I$ and its projection on the $z$-axis $K$, defined as
\begin{equation}
\begin{split}
\Psi^{IK}(\vec{r}_1,\vec{r}_2,\xi)
&=\sum_\alpha c_{0,\alpha}\left[\Theta^{J}_\alpha(\vec{r}_1,\vec{r}_2)\,\Phi^{0}_0(\xi)\right]^{IK}\\
&+\sum_{\alpha,\nu}c_{1,\alpha\nu} \left[\Theta^{J}_\alpha(\vec{r}_1,\vec{r}_2)\,\Phi^{L_\nu}_\nu(\xi)\right]^{IK},
\end{split}
\label{wf}
\end{equation}
where $\Theta^{JM}_{\alpha}$ is the uncorrelated two-body wave function of valence nucleons coupled with angular momentum $J$ and its projection on the $z$-axis $M$, and $\Psi^{L_\nu M_\nu}_\nu(\xi)$ is the core intrinsic wave function with multipolarity $L_\nu$ and its projection on the $z$-axis $M_\nu$. The subscript $\alpha$ means the set of good quantum numbers of particles $1$ and $2$, namely $(n_1,j_1,l_1)$ and $(n_2,j_2,l_2)$. The two-body wave function, $\Theta^{JM}_\alpha$, is given in an anti-symmetrized form of two single-particle wave functions of $\psi_{njl}$. The variable $\xi$ indicates the core intrinsic coordinate. The first term of Eq.~\eqref{wf} is the superposition of the two-body wave function and the core ground state ($\Phi^{00}_0$), and the second term is that of the two-body wave function and core excited states ($\nu\ne0$).

The core ground state is defined as
\begin{eqnarray}
Q_\nu \Phi^{00}_0(\xi)=0,
\end{eqnarray}
and the excited states are
\begin{equation}
\Phi^{L_\nu M_\nu}_\nu(\xi)=Q_\nu^\dagger\Phi^{00}_0(\xi),
\end{equation}
where $Q_\nu^\dagger$ and $Q_\nu$ are the phonon creation and annihilation operators of core vibrational states, respectively. The function $\Phi_\nu^{L_\nu M_\nu}$ is the eigenstate of $H_{core}$,
\begin{equation}
\begin{split}
H_{core}\Psi_\nu^{L_\nu M_\nu}(\xi)&=E_\nu\Psi_\nu^{L_\nu M_\nu}(\xi),
\end{split}
\end{equation}
where $E_\nu$ is the excitation energy of core nucleus. The transition density operator of Eq.~\eqref{funcderiv},~\eqref{residualISIV},~\eqref{3bodyH2} is expressed by the phonon creation and annihilation operators~\cite{BohrMottelson};
\begin{equation}
\delta\hat{\rho}_\nu^\beta(\vec{r})=\delta\rho_\nu^\beta(\vec{r})Q^\dagger_\nu+\delta\rho_\nu^{\beta*}(\vec{r})Q_\nu,
\label{traden2}
\end{equation}
where $\delta\rho_\nu^\beta(\vec{r})$ is the transition density of $\nu$ state.

The zero-range type interaction is used as the pairing interaction of $V_p(\vec{r}_1,\vec{r}_2)$ in this work. It is separated into the spin-singlet and spin-triplet components as~\cite{Tanimura2014}
\begin{equation}
\begin{split}
&V_{p}(\vec{r}_1,\vec{r}_2)= \delta(\vec{r}_1-\vec{r}_2)\\
&\times\left\{ \hat{P}_{s}v_s\Big(1+x_s f^{\alpha_s}(r_1)\Big)
+\hat{P}_{t}v_t\Big(1+x_t f^{\alpha_t}(r_1)\Big) \right\},
\end{split}
\label{NNforce}
\end{equation}
where $f(r)=(1+\exp[(r-R_n)/a])^{-1}$. The operators $\hat{P}_s$ and $\hat{P}_t$ project the uncorrelated two-body wave function on the spin-singlet and spin-triplet channels, respectively. The parameters $v_s$, $x_s$, $\alpha_s$, $v_t$, $x_t$, $\alpha_t$, $R_n$, and $a$ are determined from experimental data, the details of which are presented in the next section.

As in Refs.~\cite{BohrMottelson, SagawaBrown1984, Hamamoto1996, Yoshida2009}, this work includes a polarization effect, but does not take into account a correlation effect which appears when the correlated ground state of the core nucleus is properly included~\cite{Mahaux1985, Bernard1979, Bernard1980}. However, the correlation effect is weaker than the polarization effect for the case of particle-core coupling~\cite{Bernard1980}.

\subsection{Model space}
\label{theory2}

\subsubsection{Vibrational states}
\label{vibrational.states}
The vibrational excited states of the core nucleus are calculated with the Skyrme random-phase-approximation (RPA), in which all of the residual interactions except the spin-orbit force are taken into account. Continuum states are discretized by introducing a box boundary condition of $r_{max}=30$ fm (a step size is equal to $0.1$ fm). Cutoff energies of single particle levels and unperturbed 1p-1h states are set to be 200 MeV. We used the SkM$^*$ effective force~\cite{SkM*}, which provides a reasonable value for the low-lying $2^+$ state \cite{Terasaki2008} (we have also tested with other effective forces, but the results of the three-body system did not change significantly). We do not include the isoscalar dipole states because it is nothing but an unphysical translational mode. The phonon states to be coupled with the valence nucleon(s) are restricted to natural parity states from $L=0$ to $5$ below $E_\nu \le 30$ MeV with the fraction of the total isoscalar or isovector strength being larger than 5\%. 
The strengths obtained exhaust the energy weighted sum-rule above $99\%$ for $L=0$ to $5$; contributions of unnatural parity states and from $L\geq6$ are negligibly small.

In RPA, the transition densities of Eq.~\eqref{traden2} are calculated by
\begin{equation}
\begin{split}
&\delta\rho_{\nu,q}(\vec{r})=\frac{1}{\sqrt{2L_\nu+1}}
\sum_{mi\in q} \left(X_{mi}^\nu+(-1)^{L_\nu}Y_{mi}^\nu\right) \\
&\times u_{n_mj_ml_m}(r) u_{n_ij_il_i}(r) \langle j_ml_m||Y_{L_\nu}|| j_il_i \rangle Y_{L_\nu M_\nu}^*(\hat{r}).
\end{split}
\label{transitiondensity}
\end{equation}
where $u_{njl}(r)$ is the radial part of the single particle wave function of $\psi_{njl}(\vec{r})$, $q$ takes the proton or neutron, and indices $m$ and $i$ specify particle and hole states, respectively. The coefficients $X_{mi}$ and $Y_{mi}$ in Eq.~\eqref{transitiondensity} are the RPA forward and backward amplitudes, respectively~\cite{RingandSchuck}. The isoscalar and isovector transition densities from Eq.~\eqref{traden2} are given by $\delta\rho^{IS}_\nu=\delta\rho_{\nu,n}+\delta\rho_{\nu,p}$ and $\delta\rho^{IV}_\nu=\delta\rho_{\nu,n}-\delta\rho_{\nu,p}$, respectively.

\subsubsection{Two-body wave functions and interactions}
\label{2body.interaction}
Nucleon mass in Eq.~\eqref{3bodyH2} is set to $m=m_n=m_p=938$ MeV/$c^2$. The core nucleus is assumed to be a spherical even-even nucleus. The core-valence nucleon interaction in the static density, $V_{ic}^0(\vec{r}_i)$, is replaced by a phenomenological Woods-Saxon potential of the same form as Eq.~(3) of ~\cite{Tanimura2014}. The spatial parameters for the Woods-Saxon potential are $R_n=1.27A^{1/3}_C$ fm, and $a=0.67$ fm. The Coulomb potential is calculated by a uniformly charged sphere of radius $R_n$ and charge $Z_ce$, and hyperfine structure is set to be $e^2/\hbar c=1/137.036$. We include all the possible combination satisfying $\epsilon_{n_1j_1l_1}+\epsilon_{n_2j_2l_2} \le E_{cut}$ as a two-body wave function of the valence nucleons $\Theta_\alpha^{JM}(\vec{r}_1,\vec{r}_2)$, where $E_{cut}$ is set to be $20$ MeV.

\begin{table}[htbp]
\caption{Parameters of the Woods-Saxon potential and pairing interaction. The values for the case of inert core \cite{Tanimura2014} are listed for comparison with those of vibrational core (vib. core).}
\begin{tabular}{cc|rr}
\hline\hline
\multicolumn{4}{c}{Woods-Saxon potential}\\
Parameter & Nucleus & vib. core & inert core \\
\hline
$v_0$ (MeV)&$^{12}$C  & $-39.56$ & $-40.63$ \\
        &$^{16}$O & $-44.85$ & $-49.21$ \\
        &$^{28}$Si & $-44.41$ & $-47.30$ \\
        &$^{32}$S & $-45.55$ & $-46.53$ \\
        &$^{40}$Ca & $-48.57$ & $-51.79$ \\
        &$^{56}$Ni & $-48.42$ & $-50.95$ \\
$v_{ls}$ (MeV fm$^2$)& & $27.7$ & $21.6$ \\
\hline
\multicolumn{4}{c}{Pairing interaction}\\
\multicolumn{2}{c}{Parameter} & vib. core & inert core\\
\hline
\multicolumn{2}{c|}{$x_s$} & $-1.210$ & $-1.229$ \\
\multicolumn{2}{c|}{$x_t$} & $-1.350$ & $-1.417$ \\
\multicolumn{2}{c|}{$\alpha_s,\alpha_t$} & $1.221$ & $1.233$ \\
\end{tabular}
\label{vvls1}
\end{table}

The spin-orbit strength, $v_{ls}$ (see Eq.~(3) of ~\cite{Tanimura2014}), is determined so as to reproduce the $LS$ splitting between the $1d_{5/2}$ and $1d_{3/2}$ states of $^{17}$O. The potential depth, $v_0$ (see Eq.~(3) of ~\cite{Tanimura2014}), is determined from neutron separation energies in the two-body system (i.e. even-even core plus neutron). To account for the core vibration effect on the neutron separation energies, the two-body system of the effective Hamiltonian of Eq.~\eqref{3bodyH2} is solved, omitting $H_2$, $V_p(\vec{r}_1,\vec{r}_2)$, and $\frac{\delta V_{ic}}{\delta\rho}(\vec{r}_2)$. The resulting Woods-Saxon potential parameters are shown in Table~\ref{vvls1}. The potential depths in this work are shallower than those of the inert core because the vibrational coupling effect decreases the single particle energies of particle states~\cite{Bernard1980, Colo2010, Cao2014}.  Note that $x_s$ is corrected to $-1.229$ from $x_s=-1.24$ in Ref.~\cite{Tanimura2014}, which was found to be a typo.

The strength parameters of the pairing interactions of Eq.~\eqref{NNforce} are determined from the proton-neutron scattering length as~\cite{Esbensen1997}
\begin{equation}
v_{s,t}=\frac{2\pi^2\hbar^2}{m}\frac{2a_{pn}^{(s,t)}}{\pi-2a_{pn}^{(s,t)}k_{cut}},
\end{equation}
where $a_{pn}^{(s)}=-23.749$ fm and $a_{pn}^{(t)}=5.424$ fm~\cite{Koester1975} are the empirical $p$-$n$ scattering length in the spin-singlet and -triplet channels, respectively, and $k_{cut}=\sqrt{mE_{cut}/\hbar^2}$. The other parameters of the pairing interactions are determined by the $0^+$, $1^+$ and $3^+$ state of $^{18}$F nucleus as done in~\cite{Tanimura2014}. The result of the pairing parameters are summarized in Table~\ref{vvls1}.

The isoscalar and isovector residual interactions of $v_{res}^{IS, IV}(\vec{r},\vec{r}')$ can be derived from the second derivative of the Skyrme energy density with respect to densities. In order to simplify our numerical calculations, in this work, we simplified the momentum dependent terms by adopting the Landau-Migdal form~\cite{SagawaBrown1984, Hamamoto1996}, in which the corresponding parameters of $t_i$, $x_i$, $\alpha$, and $k_F$ are taken from SkM$^*$ force.

The Hamiltonian matrix of Eq.~\eqref{3bodyH2} is sparse and its dimension reaches about a few hundred thousand. The eigenvalue problem of such large sparse matrices is numerically solved by the JADAMILU code~\cite{Jacobi-Davidson}.

\section{Numerical result}
\label{result}
\subsection{Magnetic moments and spin-isospin excitations}
\label{N=Z}

\begin{table*}[htbp]
\caption{Comparison of $\Delta E=E_{1^+_1}-E_{0^+_1}$, magnetic moment $\mu_N$, and magnetic dipole transition $B$($M1$) for $N=Z$ odd-odd nuclei calculated by the three-body model with a vibrational core (vib) and inert core (inert) with experimental data (exp). The nucleus in parentheses indicates the core nucleus. Experimental data of $\mu$ for $^{14}$N and $^{58}$Cu are taken from \cite{Proctor} and \cite{Stone2008}, respectively, and those of $\Delta E$ and $B$($M1$) from the National Nuclear Data Center~\cite{nndc}. The numbers in parentheses give the experimental errors in the last digits. Errors of the experimental data of $\Delta E$ are omitted because they are very small.}
\begin{tabular}{cc|c|c|c|c|c|c}
\hline\hline
\multicolumn{2}{c|}{Observables} & $^{14}$N ($^{12}$C) & $^{18}$F ($^{16}$O) & $^{30}$P ($^{28}$Si) & $^{34}$Cl ($^{32}$S) & $^{42}$Sc ($^{40}$Ca) & $^{58}$Cu ($^{56}$Ni) \\
\hline\hline
$\Delta E$ & exp & $2.31$ & $1.04$ & $0.68$ & $-0.46$ & $-0.61 $ & $0.20$ \\
(MeV)        & inert   & $0.05$  & $1.04$ & 0.02 & $-0.69$ & $-0.61$ & 0.68 \\
                & vib & $0.48$ & $1.04$ & 0.18 & $-0.29$ & $-0.28$ & 0.08 \\
\hline
$\mu$      & exp & $0.404$ & - & - & - & - & $0.52 (8)$ \\
$(\mu_N)$ & inert & 0.379 & 0.834 & 0.318 & 0.426 & 0.686 & 0.283 \\
                & vib & 0.406 & 0.78 & 0.596 & 0.389 & 0.607 & 0.432 \\
\hline
$B$($M1$) & exp   & $0.047 (2)$ & $19.71 (3.47)$ & $1.32 (14)$ & $0.077 (6)$ & $6.16 (265)$ & -\\
$(\mu_N^2)$ & inert & 0.678 & 18.1 & 0.37 & 0.15 & 6.79 & 0.658 \\
                   & vib & 0.668 & 16.1 & 3.35 & 0.13 & 5.44 & 5.977 \\
\end{tabular}
\label{1plus}
\end{table*}

In this section, we present the following: the energy difference between $1^+_1$ and $0^+_1$ states ($\Delta E$), the magnetic moment ($\mu$) and the magnetic dipole moment ($B$($M1$)) from the first $0^+$ ($1^+$ for $^{34}$Cl and $^{42}$Sc) excited state to the $1^+$ ($0^+$) ground state, and the low-lying GT strength ($B$(GT)). One of the interests of this work is to see the variance between our results and those using three-body model with an inert core, so that our outcomes are compared with the previous work of Ref.~\cite{Tanimura2014}. 
In calculating $\mu$, $B$($M1$), and $B$(GT), we did not take into account the contributions from the core nucleus (the third term of Eq.~\eqref{amp} in Appendix) because they are expected to be small at a low energy.

The results of $\Delta E$,  $\mu$, and $B$($M1$) are listed in Table~\ref{1plus}. The three-body model calculations of vibrational core (vib) are compared with those using an inert core (inert) and experimental data (exp). The three-body model using an inert core reasonably reproduces $\Delta E$ of $^{34}$Cl and $^{42}$Sc, $\mu$ of $^{14}$N, and $B$($M1$) of $^{18}$F and $^{42}$Sc. However, the other quantities differ substantially from the experimental data. The gaps between the theoretical calculation and the experimental data are significantly closed by including the core vibration. For example, $\Delta E$ for $^{14}$N, $^{30}$P, $^{34}$Cl, and $^{58}$Cu, $\mu$ of $^{14}$N and $^{58}$Cu, and $B$($M1$) of $^{34}$Cl are improved. The magnetic dipole strength $B$($M1$) of $^{18}$F and $^{42}$Sc are also reproduced within the experimental errors. This result indicates the importance of coupling with core vibrational states. Note that $^{14}$N and $^{30}$P have a deformed core, so that the deformation effect and the coupling with rotationally excited states would provide even further improvement. A large discrepancy of the $B$($M1$) of $^{14}$N from the experimental data, already found in the three-body model including an inert core, is due to the three-body force~\cite{Maris2011,Ekstrom2014} which is difficult to include it in the present framework.

Coupling to higher order configuration causes quenching and damping of spin-isospin excitations, and the missing strengths are brought to a higher energy region or converted to a $\Delta$-hole excitation~\cite{Heyde2010, Ichimura2006}. The present three-body model also clarifies the damping of $B$($M1$) by allowing the core nuclei to vibrationally excite. We can see in Table~\ref{1plus} that the $B$($M1$) of the three-body model using a vibrational core are certainly smaller than that of the model using an inert core for $^{14}$N, $^{18}$F, $^{34}$Cl, and $^{42}$Sc. However, for $^{30}$P and $^{58}$Cu, the $B$($M1$) are enhanced rather than reduced. Those nuclei have a different configuration for the $1^+$ ground state between the three-body models of inert core and vibrational core, as discussed in sec.~\ref{sec.su4multi}.

\begin{table}[htp]
\caption{Low-lying GT strengths (in units of $g_A^2/4\pi$) for $^{18}$O$\rightarrow^{18}$F, $^{42}$Ca$\rightarrow^{42}$Sc, and $^{58}$Ni$\rightarrow^{58}$Cu transitions calculated by the three-body model with an inert core (inert) and vibrational core (vib.). The $B$(GT) of the three-body model with an inert core are taken from Ref.~\cite{Tanimura2014} and are corrected by a factor of two. Experimental data are taken from Refs.~\cite{Tilley1995,Kurtukian2009,Fujita2002}.}
\begin{tabular}{cccc|ccc}
\hline
\hline
&\multicolumn{6}{c}{$^{18}$O$(g.s.)\rightarrow^{18}$F ($1^+$)} \\
&\multicolumn{3}{c|}{$E_{1^+}$ (MeV)} & \multicolumn{3}{c}{$B$(GT) ($g_A^2/4\pi$)} \\
\#&inert & vib. & exp. & inert & vib. & exp.\\
\hline
1&$0.0$ & $0.0$ & $0.0$ & $4.96$ & $4.16$ & $3.11\pm0.03$\\
2&$4.79$ & $5.41$ & - & $0.056$ & $0.106$ & - \\
3&$6.87$ & $7.05$  & - & $0.072$ & $0.096$ & - \\
\hline\hline
&\multicolumn{6}{c}{$^{42}$Ca$(g.s.)\rightarrow^{42}$Sc ($1^+$)}\\
&\multicolumn{3}{c|}{$E_{1^+}$ (MeV)} & \multicolumn{3}{c}{$B$(GT) ($g_A^2/4\pi$)} \\
\#&inert & vib. & exp. & inert & vib. & exp.\\
\hline
1&$0.61$ & $0.28$ & $0.61$ & $3.60$ & $3.08$ & $2.16\pm0.15$\\
2&- & - & $1.89$ & - & - & $0.09\pm0.02$  \\
3&$3.71$ & $4.99$  & $3.69$ & $0.79$ & $0.57$ & $0.15\pm0.03$\\
\hline\hline
&\multicolumn{6}{c}{$^{58}$Ni$(g.s.)\rightarrow^{58}$Cu ($1^+$)} \\
&\multicolumn{3}{c|}{$E_{1^+}$ (MeV)} & \multicolumn{3}{c}{$B$(GT) ($g_A^2/4\pi$)} \\
\#&inert & vib. & exp. & inert & vib. & exp.\\
\hline
1&$0.0$ & $0.0$ & $0.0$ & $0.194$ & $1.424$ & $0.155\pm0.010$\\
2&$1.24$ & $0.92$  & $1.05$ & $1.48$ & $1.77$ & $0.30\pm0.04$\\
\hline
\end{tabular}
\label{lowGT}
\end{table}

Next, we discuss the damping and quenching of $B$(GT). The GT transitions are frequently studied in a framework beyond the standard 1 particle-1 hole RPA in order to understand the effect of higher-order configurations on quenching and damping~\cite{Drozdz1990, Niu2015, Minato2016}. Because the GT transition is a major contributor to $\beta$-decay for most nuclei, a thorough understanding of it is important for various fields, for example, nucleosynthesis in astrophysics and nuclear engineering. We study the GT transition from the $0^+$ ground state of $(N_c+2,Z_c)$ even-even nucleus to $1^+$ states of $(N_c+1,Z_c+1)$ odd-odd nucleus with the three-body model. Parent $(N_c+2,Z_c)$ nuclei, i.e. the system of core plus two neutrons, are calculated in the same manner as that of core plus proton and neutron. It should be mentioned that the studies of quenching and damping of GT strengths for even-even nuclei has been carried out with the particle-vibration coupling in a Skyrme interaction~\cite{Niu2012, Niu2014, Niu2016}. The purpose of this work is, in contrast to them, to see the effect of the SU(4) symmetry on the low-lying  GT transitions by the three-body model (the details concerning the SU(4) symmetry in the nuclei studied in this work will be discussed in sec.~\ref{sec.su4multi}).

The low-lying $1^+$ excitation energies and the $B$(GT) (in units of $g_A^2/4\pi$, where $g_A$ is the axial vector coupling constant of a free nucleon) for $^{18}$O, $^{42}$Ca, and $^{58}$Ni calculated by the three-body model are listed in Table.~\ref{lowGT} together with the experimental data. For the three-body model using an inert core, the calculated results of $B$(GT) overestimate the experimental data. For $^{18}$F ($^{42}$Sc), $B$(GT) of the first $1^+$ state exhausts $83\%$ ($60\%$) of the Ikeda sum rule $3(N-Z)=6$. By considering the core vibration, the $B$(GT) is damped and exhausts $69\%$ ($51\%$) of the sum-rule for $^{18}$F ($^{42}$Sc). As a result, the $B$(GT) gets closer to the experimental data of the first $1^+$ state. On the other hand, $B$(GT) of $^{58}$Cu are enhanced by the core vibration both for the first and second $1^+$ states, and they are one order larger than the experimental data. The enhancement from inert core and the overestimation of experimental data are discussed in sec~\ref{sec.su4multi}.

\begin{figure*}[htb]
\includegraphics[width=0.40\linewidth]{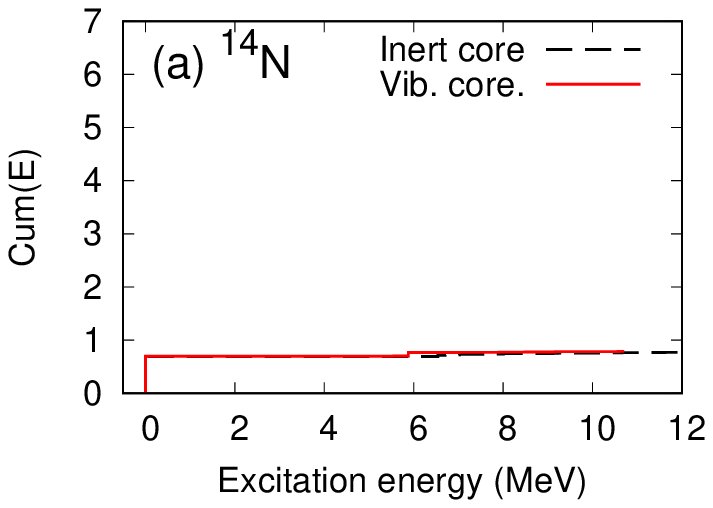}
\includegraphics[width=0.40\linewidth]{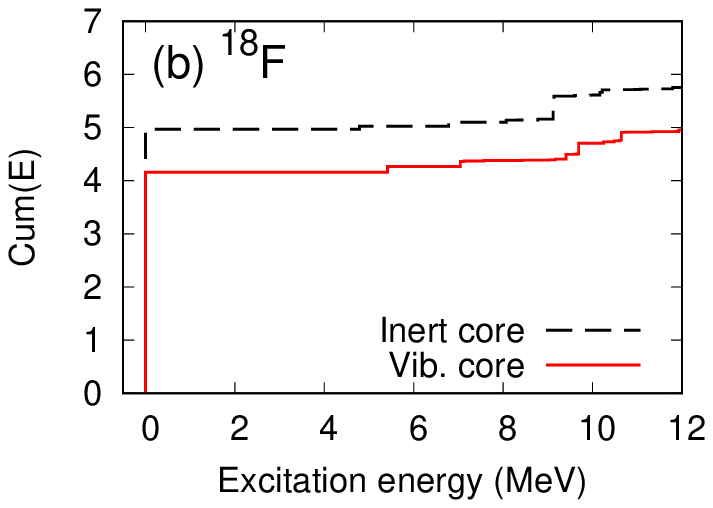}
\includegraphics[width=0.40\linewidth]{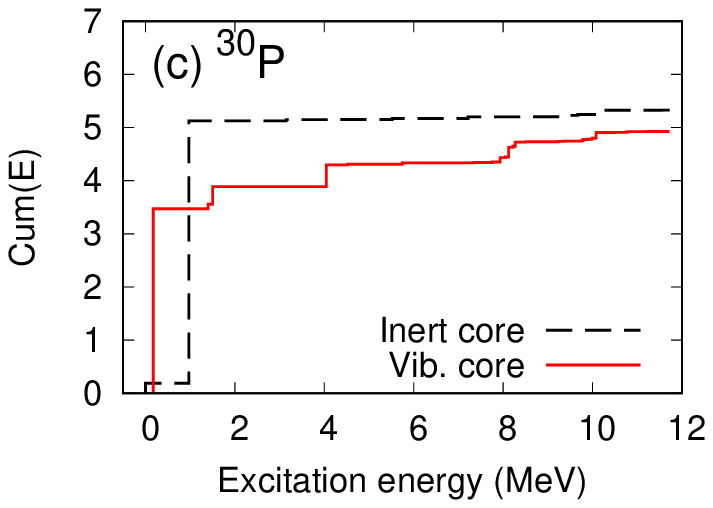}
\includegraphics[width=0.40\linewidth]{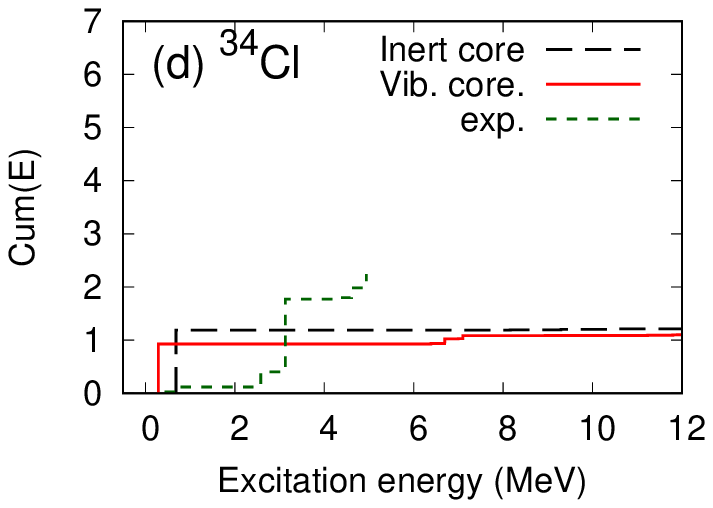}
\includegraphics[width=0.40\linewidth]{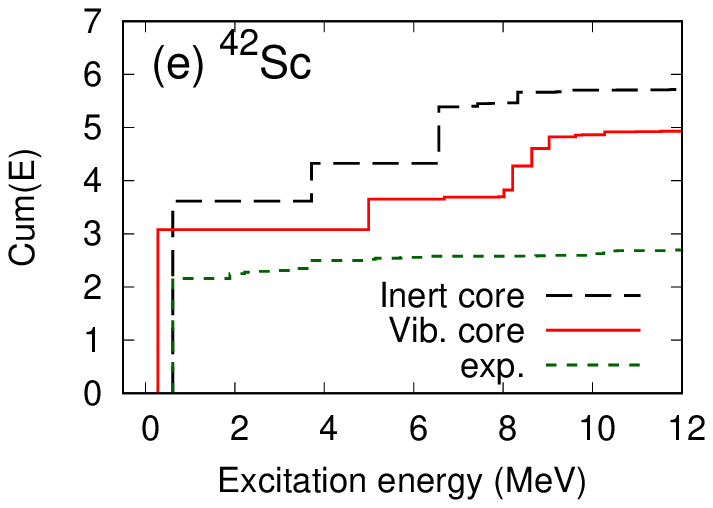}
\includegraphics[width=0.40\linewidth]{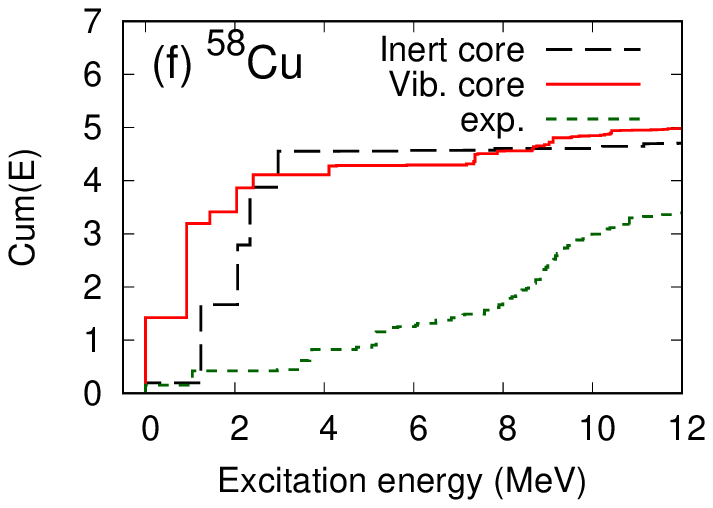}
\caption{Cumulative of $B$(GT) in units of $g_A^2/4\pi$ as a function of excitation energy calculated by the three-body model using an inert core and vibrational core. A contribution of the GT transition from the core nucleus is not included. The experimental data are taken from Ref.~\cite{Fujita2007} for $^{34}$Cl and Ref.~\cite{Fujita2014, Fujita2015} for $^{42}$Sc, and Refs.~\cite{Fujita2002, Fujita2007b} for $^{58}$Cu. No quenching factor is included in the calculation.}
\label{sumGT}
\end{figure*}

To demonstrate the quenching of $B$(GT), we plot cumulative $B$(GT) defined as
\begin{equation}
{\rm Cum}(E)=\sum_{\nu \in E}B({\rm GT},\nu)
\end{equation}
in Fig.~\ref{sumGT} as a function of excitation energy $E$ of the $1^+$ state up to 12 MeV. Note that the results shown in Fig.~\ref{sumGT} are the GT transitions between the valence nucleons and those of the core nucleus (the third term of Eq.~\eqref{amp} in Appendix) is omitted. We have checked by RPA with the SGII force~\cite{Giai1981} that first significant GT transitions of core nuclei are about 5 MeV for $^{14}$N, 11 MeV for $^{30}$P, 10 MeV for $^{58}$Cu and above 12 MeV for the other nuclei. We can observe in Fig.~\ref{sumGT} that Cum($E$) at $E=12$ MeV for $^{18}$F, $^{30}$P, $^{42}$Sc and $^{58}$Cu are approximately 6, which is a reasonable value with respect to the Ikeda sum-rule, while those of $^{14}$N and $^{34}$Cl are only about 1. The missing strengths for $^{14}$N ($^{34}$Cl) are formed by the transition of the core nucleus from $\nu1p_{3/2} (1d_{5/2})$ to $\pi1p_{1/2} (1d_{3/2})$ states and distributes around 5 MeV. For $^{58}$Cu, the giant GT strength, which is mainly formed by the transition of the core nucleus from $\nu1f_{7/2}$ to $\pi 1f_{5/2}$ states, appears around 10 MeV in the RPA calculation~\cite{Bai2014} in addition to the Cum($E$) shown in Fig~\ref{sumGT}.

From Fig.~\ref{sumGT}, the quenching of $B$(GT) can be clearly observed for $^{18}$F, $^{30}$P, and $^{42}$Sc. The ratio of Cum($E$) at $E=12$ MeV for the three-body model using a vibrational core to that using an inert core are approximately $0.80$ and is consistent with the results of particle vibration coupling~\cite{Niu2014} and the second Tamm-Dancoff-Approximation~\cite{Minato2016}. The quenching effect for $^{14}$N, $^{34}$Cl, and $^{58}$Cu is not as obvious as it is with the other nuclei. It will appear in the giant GT components coming from the core nuclei.

\begin{figure}[htb]
\includegraphics[width=0.80\linewidth]{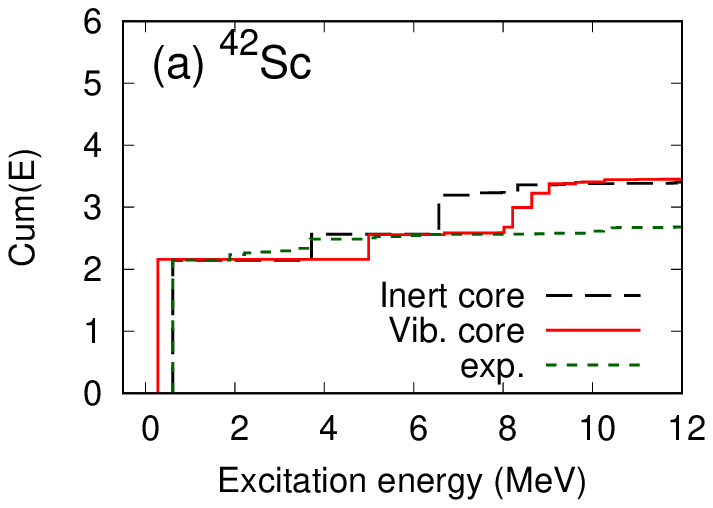}
\includegraphics[width=0.80\linewidth]{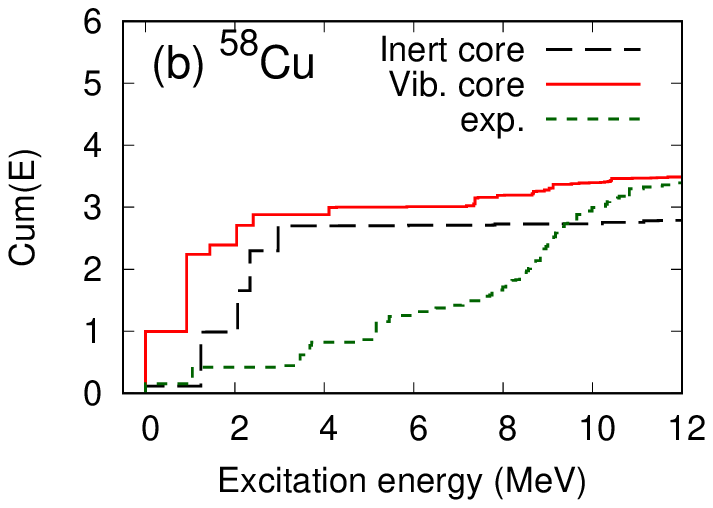}
\caption{Cumulative of $B$(GT) in units of $g_A^2/4\pi$ as a function of excitation energy for (a) $^{42}$Sc and (b) $^{58}$Cu. Experimental data are taken from Ref.~\cite{Fujita2014, Fujita2015} for $^{42}$Sc and Refs.~\cite{Fujita2002, Fujita2007b} for $^{58}$Cu. Quenching factors of $(0.77)^2$ and $(0.84)^2$ are included in the calculation of the three-body model using an inert core and vibrational core, respectively. A contribution of the GT transition from the core nucleus is not included.}
\label{sumGTb}
\end{figure}

For $^{42}$Sc and $^{58}$Cu, Cum($E$) of the three-body models overestimates the experimental data. However, for those $pf$-shell nuclei, the quenching factor has to be commonly introduced in the shell model calculation in order to reproduce the experimental data~\cite{Honma2005,Martinez-Pinedo1996}. In the present models, we determine the quenching factor so as to reproduce the experimental $B$(GT) for the first $1^+$ state of $^{42}$Sc and obtain $(0.77\pm0.03)^2$ and $(0.84\pm0.03)^2$ for the three-body model using an inert core and vibrational core, respectively.
These values are consistent with studies used in Refs.~\cite{Martinez-Pinedo1993, Nakada1996, Jokinen1998}. The result with the quenching factors is shown in Fig.~\ref{sumGTb}. We can see the three-body models reasonably reproduce the overall $B$(GT) distribution of $^{42}$Sc up to $8(6)$ MeV for the three-body model using a vibrational (an inert) core. On the other hand, the three-body models for $^{58}$Cu overestimate the $B$(GT) strengths in low energy regions. This problem may be due to the approximations used in this work and further discussion is given in the next section.

\subsection{SU(4) multiplet}
\label{sec.su4multi}

With the limit of no spin-orbit and Coulomb forces, proton and neutron states in the same orbit are degenerate and form the SU(4) multiplets~\cite{Wigner1937}. In this circumstance, the spin-isospin operator $\vec{\sigma}\,\vec{\tau}$ connects the SU(4) multiplets and the transition energies between $0^+$ and $1^+$ states appear at zero with remarkably strong $B$($M1$) and $B$(GT). Even though the spin-orbit and Coulomb forces work in nuclei, it is expected that the SU(4) symmetry appears for some nuclei. The picture of core plus valence nucleons in the three-body model is suitable for discussing the SU(4) symmetry of two valence nucleons in $N=Z$ odd-odd nuclei. It was seen in three-body model using an inert core~\cite{Tanimura2014} and other calculations~\cite{Halse1989, Vogel1993, Isacker1995} that $^{18}$F and $^{42}$Sc have a SU(4) multiplet because the $0^+$ and $1^+$ states are largely dominated by spin $S=0$ and $S=1$ components, respectively, and they show a large $B$($M1$) and $B$(GT). 
This begs the following question: what is the effect of core vibration on the SU(4) symmetry?

\begin{figure}
\includegraphics[width=0.95\linewidth]{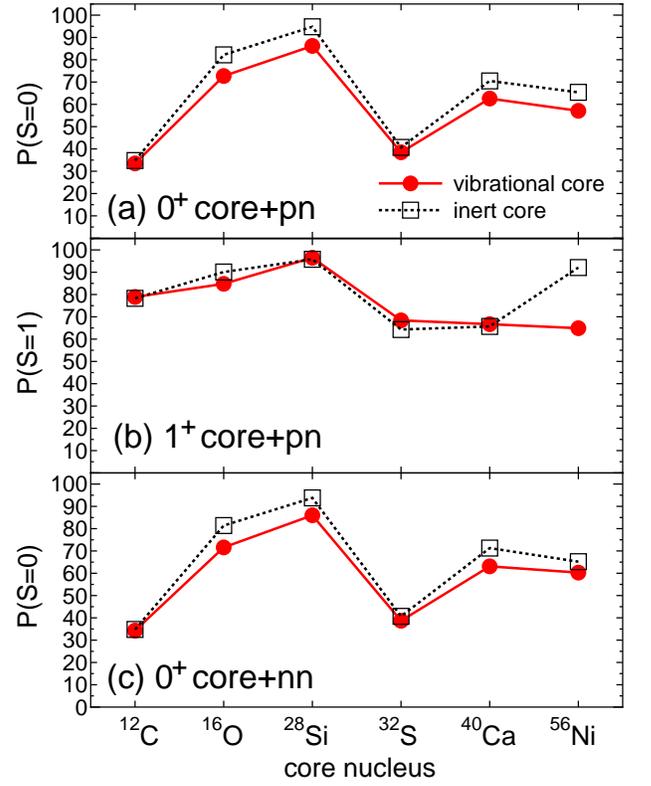}
\caption{The probabilities of (a)the spin $S=0$ component in the wave function of the $0^+$ state for the core+$pn$ system, (b)spin $S=1$ component in the wave function of the $0^+$ state for the core+$pn$ system, and (c)the spin $S=0$ component in the wave function of the $0^+$ state for the core+$nn$ system. The dotted and solid lines are the results of the three-body model using an inert core and vibrational core, respectively.}
\label{table.SU4}
\end{figure}

\begin{figure}
\includegraphics[width=0.95\linewidth]{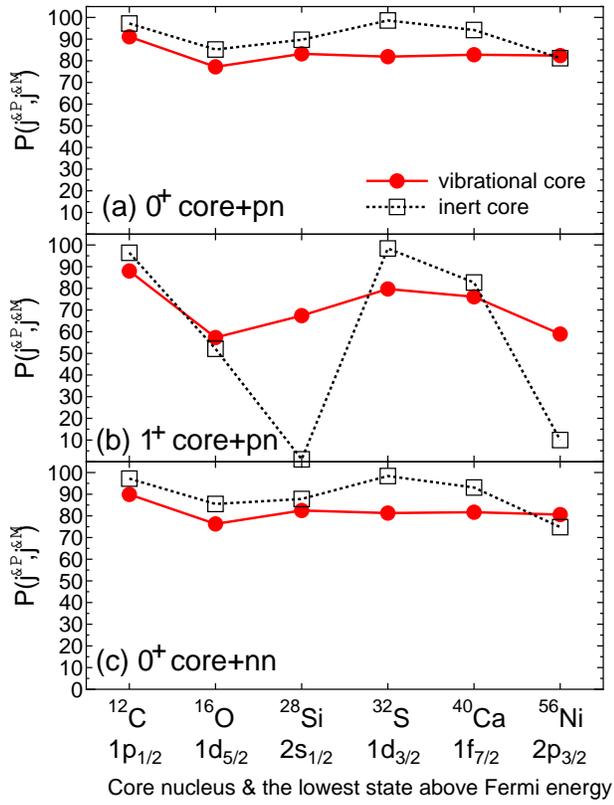}
\caption{The probabilities of particle-particle configuration in the lowest particle state, $P(j^\pi,j^\nu)$, in the wave function of $0^+$ and $1^+$ states, where $j$ is the lowest state above the Fermi energy. The dotted and solid lines are the results of the three-body model using an inert core and vibrational core, respectively.}
\label{table.SU4b}
\end{figure}

To discuss the SU(4) multiplet for $N=Z$ nuclei, it is convenient to consider the fraction of spin singlet ($S=0$) and triplet ($S=1$) components in the valence nucleon subsystem, which are defined as $P(S)$. It is calculated by using the $LS-jj$ coupling coefficient:
\begin{equation}
|(j_\pi\,j_\nu)J\rangle
=\sum_{L,S}
\left\{
\begin{tabular}{ccc}
$l_\pi$ & $l_\nu$ & $L$\\
$\frac{1}{2}$ & $\frac{1}{2}$ & $S$\\
$j_\pi$ & $j_\nu$ & $J$\\
\end{tabular}
\right\}
\hat{L}\hat{S}\hat{j}_\pi\hat{j}_\nu|(l_\pi\l_\nu)LS;J\rangle.
\end{equation}
The probability of particle-particle configuration with total angular momentum $j$ in the lowest particle state, defined as $P(j^\pi,j^\nu)=|c_{0,\alpha}|^2+\sum_\nu|c_{1,\alpha\nu}|^2$, where $\alpha \in (j^\pi,j^\nu)$, is also useful in the following discussion. The results of $P(S)$ for the first $0^+$ and $1^+$ states for core+$pn$ (i.e. $N=Z$ odd-odd nuclei), and the $0^+$ ground state for core+$nn$ (i.e. $N=Z+2$ even-even nuclei) are shown in Fig.~\ref{table.SU4}, and those of $P(j^\pi,j^\nu)$ are plotted in Fig.~\ref{table.SU4b}.

We can see in Fig.~\ref{table.SU4}(a) and (c) that $P(S=0)$ for the core+$pn$ and core+$nn$ systems calculated by the three-body model using a vibrational core are slightly smaller than those by inert core, but the variations are not large. This indicates that the core vibration has an effect disturbing the spin anti-parallel alignment of the valence nucleons, however its influence is small. Similarly, in Fig.~\ref{table.SU4}(b), $P(S=1)$ for the core+$pn$ system calculated by the three-body model of vibrational core are almost the same as those by inert core. However, we can see an exception in $^{58}$Cu.

The effect of core vibration is also limited for $P(j^\pi,j^\nu)$. In Fig.~\ref{table.SU4b}(a) and (c), $P(j^\pi,j^\nu)$ values for the $0^+$ core+$pn$ and core+$nn$ systems calculated by the three-body model using a vibrational core and inert core show almost the same results, although the vibrational core gives little smaller values than the inert core. In Fig.~\ref{table.SU4b}(b), $P(j^\pi,j^\nu)$ values for the $1^+$ core+$pn$ system show a similar result between the vibrational core and the inert core, however, $^{30}$P and $^{58}$Cu exhibit a large difference between them. We will discuss this abnormality later.

For $^{18}$F ($^{16}$O+$pn$ system) and $^{42}$Sc ($^{40}$Ca+$pn$ system), the $0^+$ and $1^+$ states are dominantly made up of $S=0$ and $S=1$ components, respectively as shown in Fig~\ref{table.SU4}, and the valence nucleons are in the same orbit ($1d_{5/2}$ for $^{18}$F and $1f_{7/2}$ for $^{42}$Sc) as displayed in Fig.~\ref{table.SU4b}. As a result, this system forms a good SU(4) symmetry in the subsystem made up of valence nucleons and a significantly large value of $B$($M1$) appears. Similarly, because the $0^+$ core+$nn$ systems are dominated by the $S=0$ component as seen in Fig.~\ref{table.SU4}(c) and the valence neutrons stay the same orbit as the $1^+$ core+$pn$ system as displayed in Fig.~\ref{table.SU4b}(b) and (c), large $B$(GT) appear for the $^{18}$O$\rightarrow^{18}$F and $^{42}$Ca$\rightarrow^{42}$Sc transitions, as seen in Table~\ref{lowGT}. As for the $^{12}$C+$pn$ and $^{32}$S+$pn$ systems, the valence nucleons occupy the same orbit between different systems as displayed in Fig.~\ref{table.SU4b}. However, $P(S=0)$ values for the $0^+$ core+$pn$ system shown in Fig.~\ref{table.SU4}(a) are small. As a result, $B$($M1$) at a low energy are suppressed for those nuclei. Similarly, because $P(S=0)$ of $0^+$ core+$nn$ system exhibit small values, $B$(GT) at a low energy are suppressed for those nuclei.

On the other hand, a small $B$($M1$) of $^{30}$P ($^{28}$Si+$pn$ system) and $^{58}$Cu ($^{56}$Ni+$pn$ system) is attributed to the fact that the valence nucleons occupy a different orbit in $1^+$ core+$pn$ system~\cite{Tanimura2014}. We can see that $P(j^\pi,j^\nu)$ values for the $1^+$ $^{28}$Si+$pn$ and $^{56}$Ni+$pn$ systems are small in case of the inert core approximation, as seen in Fig.~\ref{table.SU4b}(b). The actual major particle-particle configurations of the $1^+$ $^{28}$Si+$pn$ and $^{56}$Ni+$pn$ systems are ($1d_{3/2},2s_{1/2}$) and $(2p_{3/2},1f_{5/2})$, respectively. The situation is changed by including core vibration. In the three-body model using a vibrational core, $P(2p_{1/2},2p_{1/2})$ values for the $1^+$ $^{28}$Si+$pn$ system and $P(2p_{3/2},2p_{3/2})$ values for the $1^+$ $^{56}$Ni+$pn$ system increase by $67\%$ and $59\%$, respectively.

\begin{figure}[htp]
\includegraphics[width=0.95\linewidth]{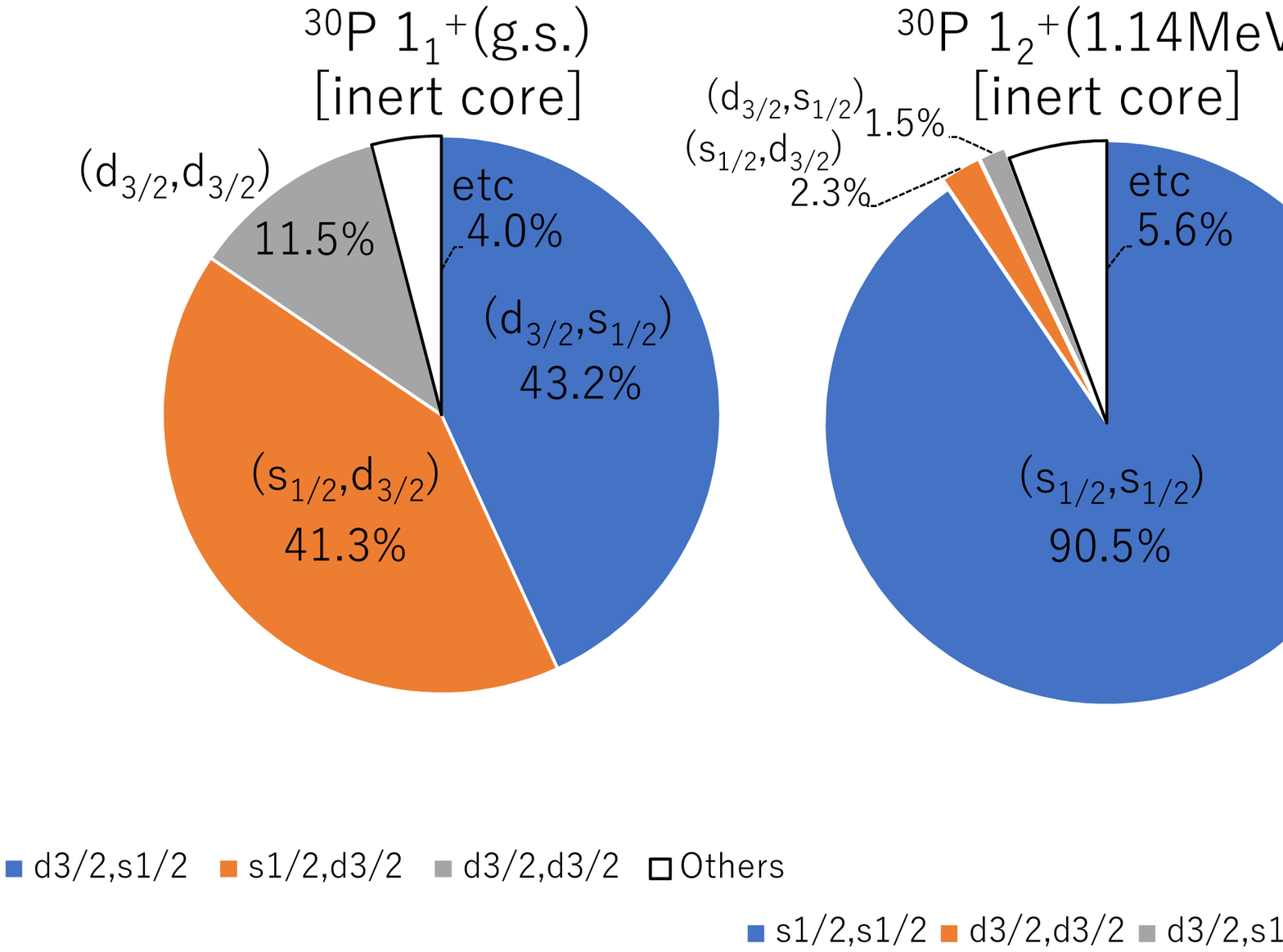}
\includegraphics[width=0.95\linewidth]{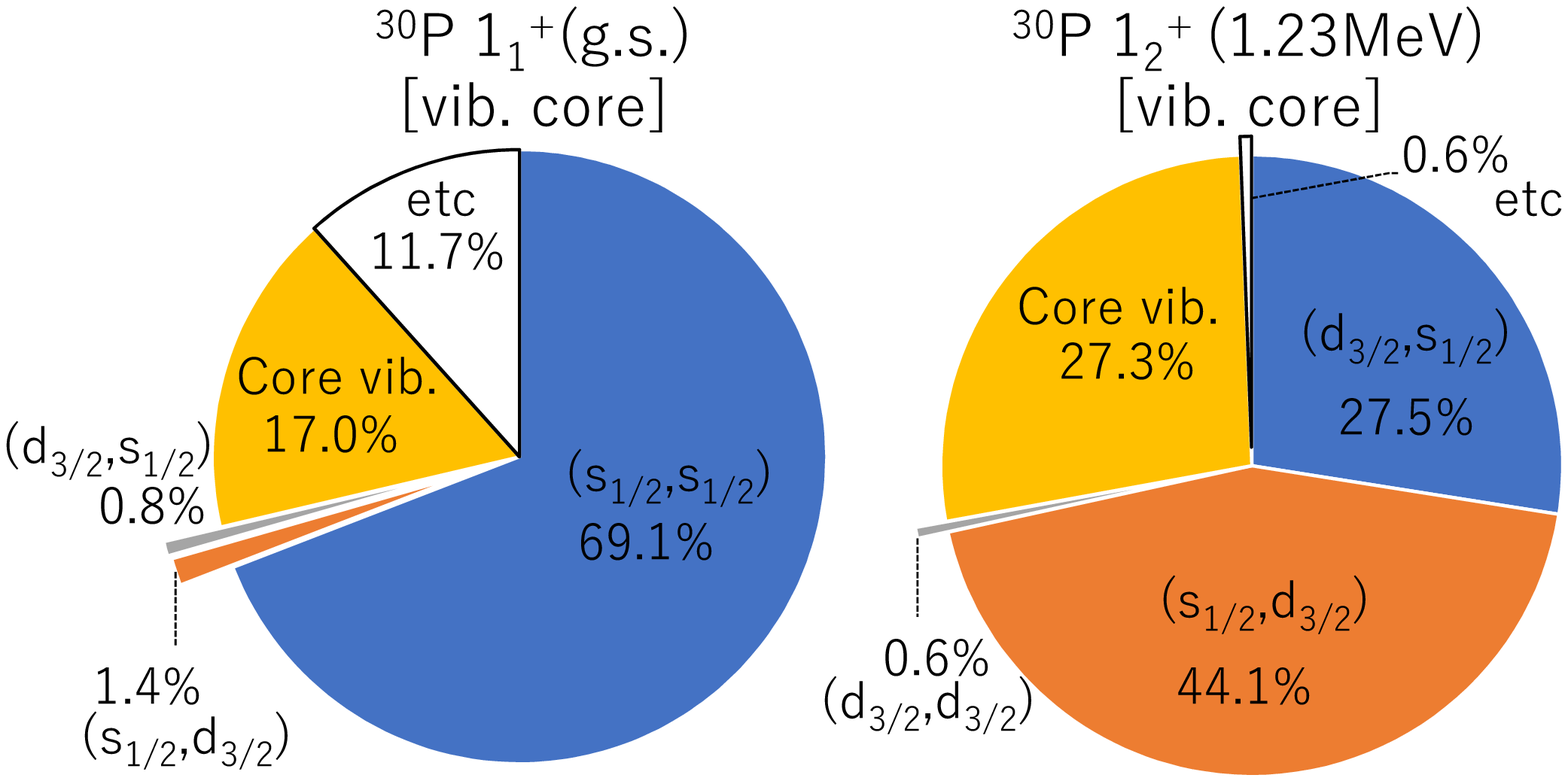}
\caption{Configurations of the ground state (the first $1^+$) and the second $1^+$ state for $^{30}$P calculated by the three-body model using a inert core and vibrational core. The upper (bottom) left and right are the results of the ground state and the second $1^+$ state in the case of the three-body model using an inert core (vibrational core), respectively.}
\label{conf.Si28}
\end{figure}

The changes of the configuration in the $1^+$ $^{28}$Si+$pn$ and $^{56}$Ni+$pn$ systems can be understood by an interchange of the ground state and the other $1^+$ excited state. To explain it, we plot the probabilities of configuration consisting of the ground state (the first $1^+$ state) and the second $1^+$ state for $^{30}$P in Fig~\ref{conf.Si28}. In the case of an inert core, the ground state (the upper left) is mainly made up of the ($d_{3/2},s_{1/2}$) and ($s_{1/2},d_{3/2}$) states with a total probability 74.0\%. The second $1^+$ state appearing at 0.97 MeV (the upper right) is mainly composed of the $(s_{1/2}, s_{1/2})$ state with 89.0\%. In the vibrational core case, the ground state (the bottom left) is mainly made up of  the ($s_{1/2},s_{1/2}$) state with 69.1\%.  The second $1^+$ state (the bottom right) is mainly the ($d_{3/2},s_{1/2}$) and ($s_{1/2},d_{3/2}$) states with a total probability 71.6\%. The configuration of the ground state (the second $1^+$ state) in the case of an inert core resembles the second $1^+$ state (the ground state) in the case of core vibration. It is qualitatively understood that the ground state configuration and the second $1^+$ state configuration are inversed by the effect of core vibration.

\begin{figure}[htp]
\includegraphics[width=0.95\linewidth]{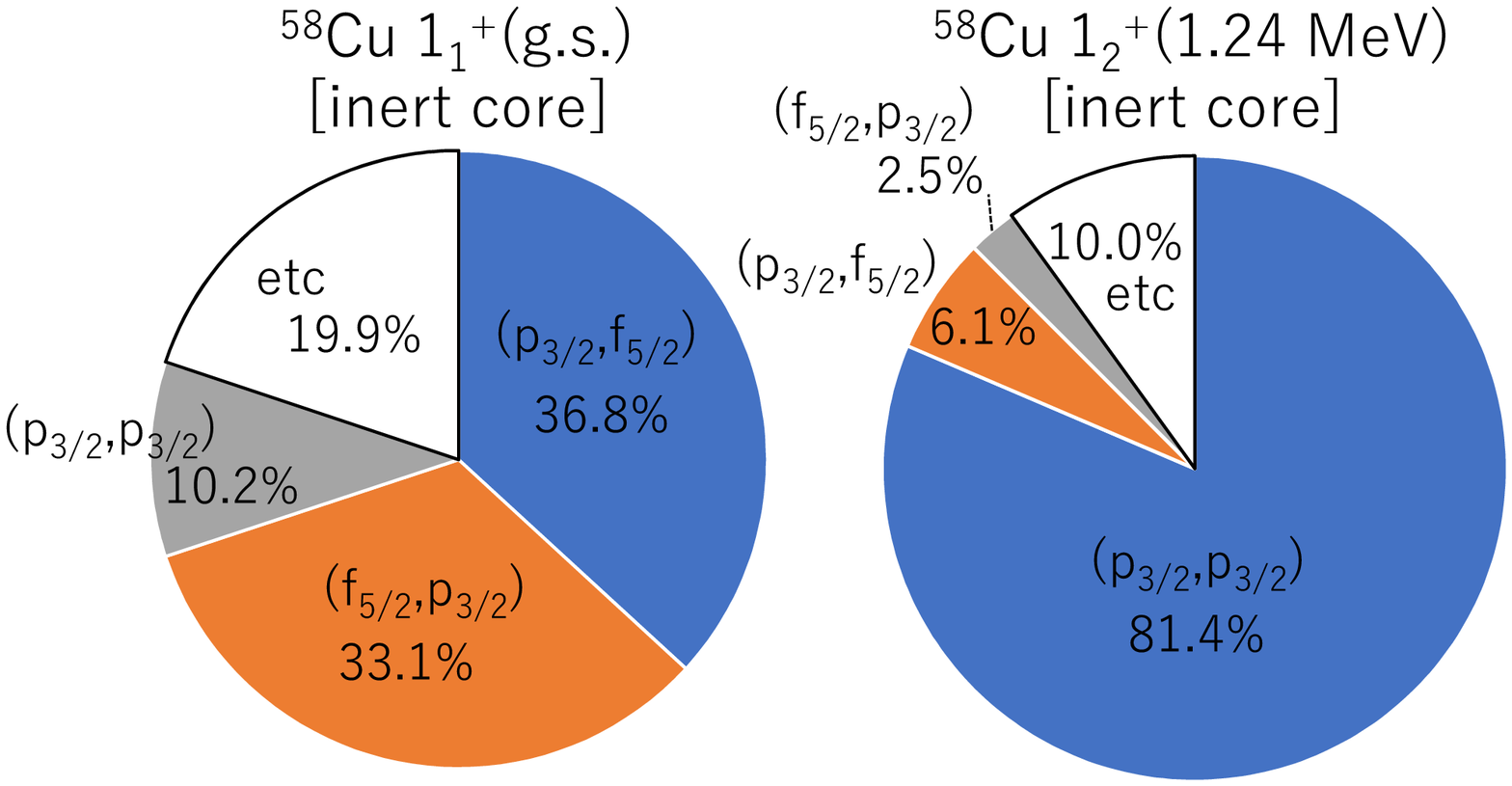}
\includegraphics[width=0.95\linewidth]{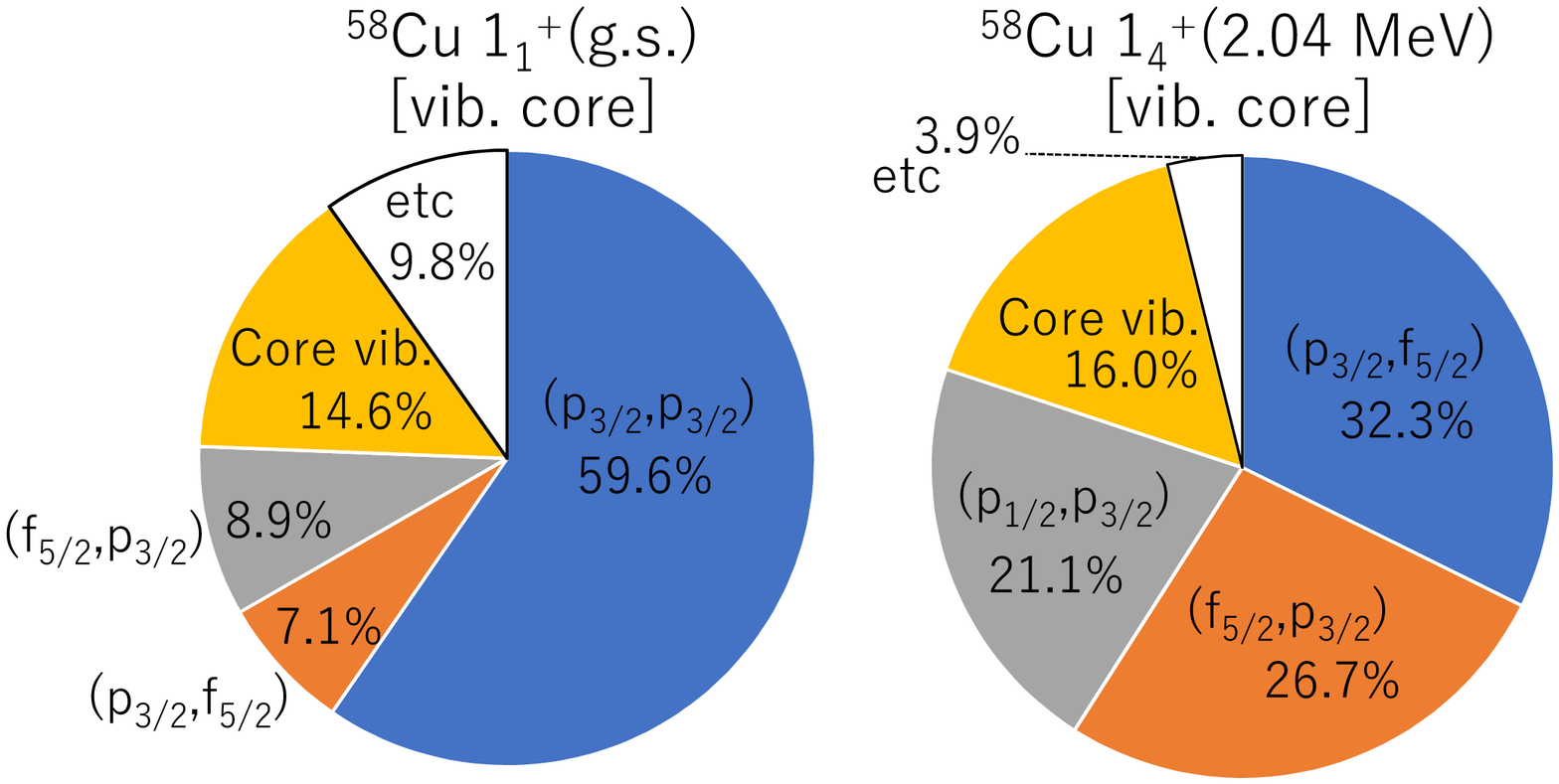}
\caption{Configurations of the ground state (the first $1^+$) and excited $1^+$ states for $^{58}$Cu calculated by the three-body model using an inert core and vibrational core. The upper left and right panels are the results of the ground state and the second $1^+$ state for the three-body model using an inert core, respectively. The bottom left and right panels are the results of the ground state and the fourth $1^+$ state for the three-body model using a vibrational core, respectively.}
\label{conf.Ni56}
\end{figure}

The configurations of the ground state and excited $1^+$ state for $^{58}$Cu are shown in Fig.~\ref{conf.Ni56}. For the inert core, the ground state (the upper left) is made up of the ($p_{3/2},f_{5/2}$) (36.5\%) and ($f_{5/2},p_{3/2}$) (32.7\%) states, and the second $1^+$ state (the upper right) is mainly made up of the $(p_{3/2}, p_{3/2})$ state (81.4\%). For the vibrational core, the ground state (the bottom left) is mainly made up of the ($p_{3/2},p_{3/2}$) state (59.6\%), and the fourth $1^+$ state (the bottom right) is made up of the ($p_{3/2},f_{5/2}$) (32.3\%) and ($f_{5/2},p_{3/2}$) states (26.7\%). As seen in $^{30}$P, the configuration of the ground state and the excited $1^+$ state looks interchanged.


Because the experimental data of $B$($M1$) for $^{30}$P has a relatively large $B$($M1$) of $1.32\pm0.14 (\mu_N^2)$, the SU(4) symmetry between $0^+$ and $1^+$ states might emerge weakly. The three-body model using an inert core underestimates the $B$($M1$), while that using a vibrational core overestimates it. Therefore, it is considered that the true probability of the particle-particle configuration $P(2s_{1/2},2s_{1/2})$ will be between the results of the three-body model using an inert core and vibrational core. Since the core nucleus $^{28}$Si is a deformed nucleus, it is expected that the $B$($M1$) is improved by including the deformation effect in addition to the vibration effect. In case of $^{58}$Cu, the experimental data of the first two $B$(GT) are small (0.455 in total) as shown in Table~\ref{lowGT}, so that the SU(4) symmetry will not be formed. However, the three-body model using a vibrational core as well as inert core shows larger $B$(GT) than the experimental data at this low energy region. We carried out a calculation for $^{58}$Cu with alternative parameters optimized to the neutron separation energy of $^{57}$Ni and $1^+$ and $0^+$ states of $^{58}$Cu, but the $B$(GT) was not reproduced. We suppose that there are several reasons why we could not reproduce the result: 
(a) the momentum dependent terms of the residual core-nucleus interaction are approximated by the Landau-Migdal form, 
(b) the tensor force between the valence particles as well as the core-nucleus is ignored. It may play a role because the tensor force changes the relative orbital angular momentum of two scattering particles by $\Delta L=2$ and disturbs the SU(4) symmetry. Using the zero-range force in the pairing interaction might also be inappropriate, and (c) the subsystem $^{57}$Cu (core plus proton) is a weakly bound system ($S_p=690$ keV), so that the exact treatment of the continuum states might be also important. 
Further study of this will be done in future work.

\subsection{Core contribution}
\label{core.contribution}
\begin{figure}
\includegraphics[width=0.95\linewidth]{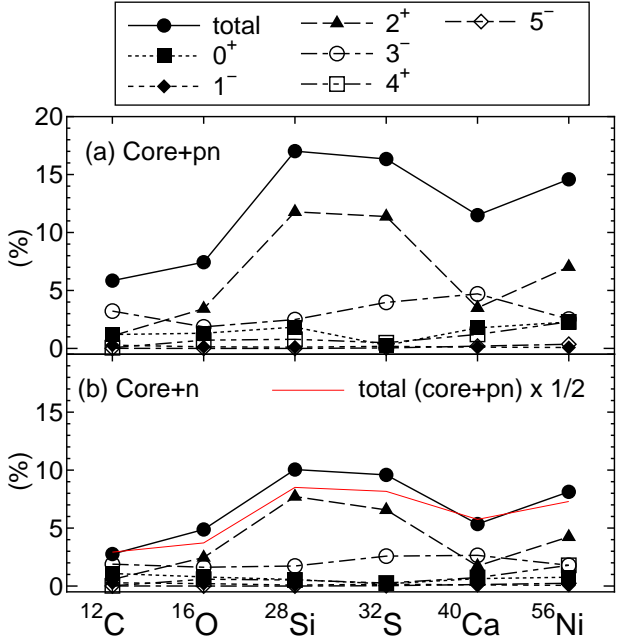}
\caption{Probabilities of core vibration consisting of the ground state of the core+$pn$ (a) and core+$n$ (b) systems are shown for core nuclei $^{12}$C, $^{16}$O, $^{28}$Si, $^{32}$S, $^{40}$Ca, and $^{56}$Ni. The thin (red) line in the panel (b) is the result of total core contribution of the core+$pn$ system multiplied by a factor of 0.5. The line is the purpose for guiding eyes.}
\label{fig.corecontri}
\end{figure}

In the previous sections, we have seen that the core vibration has a considerable influence on spin-isospin properties. Now let's see how much the core vibration contributes to the wave function of Eq.~\eqref{wf}. In Fig.~\ref{fig.corecontri}, the probabilities of core vibration consisting of the ground state of the (a) core+$pn$ and (b) core+$n$ systems are shown. We can see that the main core contribution is from the $2^+$ or $3^-$ states, while the contribution from the $1^-$ and $5^-$ states are negligibly small. The total of the core contributions of the core+$pn$ system for $^{28}$Si, $^{32}$S, $^{40}$Ca, and $^{56}$Ni are about 15 \%. In particular, the $2^+$ state contributes significantly for $^{28}$Si and $^{32}$S, because of the low-lying strong isoscalar $2^+$ states appearing at 2.4 MeV and 3.1 MeV, respectively. For the core+$pn$ systems of $^{12}$C and $^{16}$O, the total of the core contributions is only about 6 and 7 \%, respectively, because their first $2^+$ and $3^-$ states calculated by RPA are high and the transition strengths are weak. However, the effect of core vibration cannot be ignored in either the ground state or spin-isospin excitations as we have seen in the previous sections.

The result of the core+$n$ system is shown in Fig.\ref{fig.corecontri}~(b). We can see that the core contributions are smaller than the core+$pn$ system because the polarization effect is caused by only one neutron. However, the contributions of each vibrational state look similar to the core+$pn$ system. We plot the total core contribution of the core+$pn$ system divided by a factor of 2 in the same panel, which is indicated by the thin line. The result looks similar to the total core contribution of the core+$n$ system. It indicates that the core polarization is almost proportional to the number of nucleons, at least up to two particles.

\subsection{Mean value of spatial distribution of valence particles}
In the Hartree-Fock (HF) limit in which there is neither pairing nor residual core-nucleus forces, the valence nucleons stay in the first lowest orbitals above the Fermi surface. If there exists a pairing interaction, the valence nucleons collide with each other and scatter into other orbitals; this is exactly what the three-body model using an inert core can calculate. If the residual core-nucleon interaction generated by core vibration exists in addition, it also brings the valence particle to other orbitals. As a result, the geometrical structure of valence nucleons will change. To investigate this, the root mean square (rms) distance between the core nucleus and the center-of-mass of two valence nucleons, $\langle r_{C-pn}^2 \rangle^{1/2}$ and that between proton and neutron, $\langle r_{pn}^2 \rangle^{1/2}$ are calculated, which are defined as
\begin{equation}
r_{C-pn}^2=\frac{1}{2I+1}\sum_K\langle \Psi^{IK}|\frac{(\vec{r}_1+\vec{r}_2)^2}{4}|\Psi^{IK}\rangle,
\end{equation}
and
\begin{equation}
r_{pn}^2=\frac{1}{2I+1}\sum_K\langle \Psi^{IK}|(\vec{r}_1-\vec{r}_2)^2|\Psi^{IK}\rangle,
\end{equation}
respectively.

\begin{figure}
\includegraphics[width=0.95\linewidth]{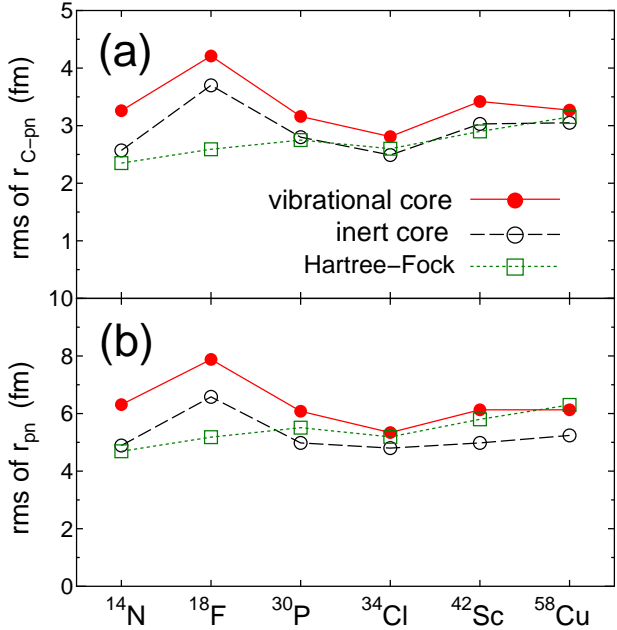}
\caption{Root mean square (rms) of geometrical variables characterizing the positions of valence nucleons and core nucleus: (a) rms distance between valence core nucleus and the center of mass of proton and neutron, $\langle r_{C-pn}^2\rangle^{1/2}$ and (b) rms distance between proton and neutron, $\langle r_{pn}^2 \rangle^{1/2}$. 
The solid, dashed and dotted lines indicate the results of the three-body model using an inert core and vibrational core, and the Hartree-Fock limit, respectively.}
\label{geo}
\end{figure}

The results of the rms of $r_{C-pn}$ for the three-body model and the HF limit are shown in Fig.~\ref{geo}(a). The three-body model using an inert core and the HF limit are almost identical except for $^{18}$F. The large $\langle r_{C-pn}\rangle^{1/2}$ of $^{18}$F is due to the scattering of the valence nucleons to the bound $s$-wave state, namely the $2s_{1/2}$ state ($\varepsilon_\pi=-0.65$ and $\varepsilon_\nu=-3.27$ MeV) from $2d_{5/2}$ state by the pairing interaction. Namely, the pairing force does not affect significantly the distance between the center-of-mass of proton and neutron and the core nucleus unless they are scattered into the $s$-wave state. If the core vibration is considered, the rms of $\langle r_{C-pn} \rangle^{1/2}$ become systematically large. Compared with the inert core, $\langle r_{C-pn}\rangle^{1/2}$ increase by $0.7$, $0.5$, $0.4$, $0.3$, $0.4$, and $0.2$ fm for $^{14}$N, $^{18}$F, $^{30}$P, $^{34}$Cl, $^{42}$Sc, and $^{58}$Cu, respectively.

Figure~\ref{geo}(b) shows the result of the rms of $\langle r_{pn}^2 \rangle^{1/2}$. The values from the three-body model using an inert core are slightly smaller than those of the HF limit with the exception of $^{14}$N and $^{18}$F. It is intuitively understood that the reductions of the rms of $\langle r_{pn}^2 \rangle^{1/2}$ occur due to the pairing interaction between the proton and neutron. A large $\langle r_{pn}^2 \rangle^{1/2}$ observed in $^{18}$F is due to the scattering to the $s$-wave state as we have seen in $\langle r_{C-pn}^2 \rangle^{1/2}$. For the case of $^{14}$N, the main ground state configuration is ($1p_{3/2},1p_{3/2}$) state (96.4\%), so that the $\langle r_{pn}^2\rangle$ gives almost the same value as that of HF limit. However, comparing the results of $\langle r_{pn}^2\rangle^{1/2}$ calculated with and without continuum states in the ground state wave function, we confirmed that the coupling with the continuum states results in little enhancement of $\langle r_{pn}^2\rangle^{1/2}$. When core vibration is considered, $\langle r_{pn}^2 \rangle^{1/2}$ systematically increase by $1.4$, $1.3$, $1.1$, $0.5$, $1.2$, and $0.9$ fm as compared to the inert core for $^{14}$N, $^{18}$F, and $^{30}$P, $^{34}$Cl, $^{42}$Sc, and $^{58}$Cu, respectively.

\section{Summary}
\label{summary}
We have studied the spin-isospin properties with the three body model including a vibrational core. The nuclei studied in this work were $N=Z$ odd-odd nuclei which consist of an even-even core nucleus plus proton and neutron. The effect of core vibration improved the results of the energy difference and $B$($M1$) between the first $0^+$ and $1^+$ states and the magnetic moments in the ground state. The damping of the low-lying $B$(GT) was observed in the three-body model using a vibrational core and the result was close to the experimental data. We also studied the $B$($M1$) and $B$(GT) from the point of view of the SU(4) symmetry.

For $^{18}$F and $^{42}$Sc, the SU(4) symmetry appears to be effective and strong $B$($M1$) and $B$(GT) are obtained at a low energy. 
Even when core excitation is considered, the situation remained almost the same. 
For $^{30}$P and $^{58}$Cu, it was found that the ground state of the first $1^+$ state and excited $1^+$ state were interchanged by the effect of core vibration. As a result, the SU(4) symmetries, which were hindered in case of inert core, turned out to have an impact, and induced significant changes in the $B$($M1$) and $B$(GT). However, the present result was not consistent with experimental data. Several issues in the present calculation of $^{30}$P and $^{58}$Cu were discussed. Further investigation considering a more general expression of the pairing and the core-nucleon forces, the core deformation, and the exact treatment of continuum states is our next perspective.

The core contribution for the ground state configurations were about 15\% for $^{30}$P, $^{34}$Cl, $^{42}$Sc, and $^{58}$Cu and a significant effect of the core vibration was observed in those nuclei. While the core contributions in the ground state wave function for $^{14}$N and $^{18}$F were only 6 to 7\%, the effect of core vibration for those nuclei cannot also be ignored in either the ground state or spin-isospin excitations. The core excitation is thus important for a detailed understanding of nuclear structure. We also found that the core contribution of the two-body system, namely core plus neutron system, was approximately half that of the three-body system.

We discussed that $\langle r_{C-pn} \rangle^{1/2}$ and $\langle r_{pn} \rangle^{1/2}$ systematically increased if core vibration was considered. This was due to the fact that the residual core-nucleon interactions induce the valence nucleons to scatter to higher orbits in addition to the pairing interaction.

Experimental study of the $B$($M1$) for $^{58}$Cu and the magnetic moment of $^{30}$P, which have not been measured, is helpful for further interpretation of the SU(4) multiplets for those nuclei. Because the $B$($M1$) for $^{18}$F and $^{42}$Sc have large uncertainties, an accurate measurement will be also important in understanding the effect of core contribution and SU(4) multiplets sufficiently.

 \begin{acknowledgments}
The author thanks W. Nazarewicz for valuable discussions and comments and Y. Fujita for providing information about experimental data and instructive comments. They also thank E. Olsen for checking manuscript throughout. F. M. thanks F. Nunes, S. Wang, and R.M. ld Betan for useful discussions about the three-body model. F. M. also acknowledges Study Abroad Program of Japan Atomic Energy Agency, and Y. T. acknowledges the financial support from the Graduate Program on Physics for the Universe of Tohoku University.
 \end{acknowledgments}

\appendix

\section{Transition amplitude}
\label{secamp}

A one-body operator $\mathcal{O}$ can be separated into the valence particles and core nucleus as,
\begin{equation}
\mathcal{O}=\mathcal{O}_{1}+\mathcal{O}_{2}+\mathcal{O}_{\xi}
\end{equation}

The reduced transition matrix of the one-body operator is calculated as using Eq.~\eqref{wf},

\begin{equation}
\begin{split}
&\langle\Psi^{I'}||\mathcal{O}||\Psi^I\rangle
=\langle\Psi^{I'}||\mathcal{O}_{1}+\mathcal{O}_{2}+{O}_{\xi}||\Psi^I\rangle\\
=&\sum_{\alpha,\alpha'}c_{0,\alpha'}^*c_{0,\alpha}
\langle \Theta_{\alpha'}^{I'}||\mathcal{O}_{1}+\mathcal{O}_{2}||\Theta_\alpha^{I}  \rangle \\
+&\sum_{\alpha,\alpha',\nu}c_{1,\alpha'\nu}^*c_{1,\alpha\nu}
\langle \left[\Theta_{\alpha'}^{J'}\otimes\Phi_c^{L_\nu}\right]^{I'}||\mathcal{O}_{1}+\mathcal{O}_{2}||\left[\Theta_\alpha^{J}\otimes\Phi_c^{L_\nu}\right]^{I}  \rangle\\
+&\sum_{\alpha,\nu,\nu'}c_{1,\alpha\nu'}^*c_{1,\alpha\nu}
\langle \left[\Theta_{\alpha}^{J}\otimes\Phi_c^{L_{\nu'}}\right]^{I'}||\mathcal{O}_{\xi}||\left[\Theta_\alpha^{J}\otimes\Phi_c^{L_\nu}\right]^{I}  \rangle \\
\end{split}
\label{amp0}
\end{equation}
Using some formulas concerning angular momentum, Eq.~\eqref{amp0} becomes
\begin{equation}
\begin{split}
&\langle\Psi^{I'}||\mathcal{O}||\Psi^I\rangle
=\sum_{\alpha,\alpha'}c_{0,\alpha'}^*c_{0,\alpha}
\langle \Theta_{\alpha'}^{I'}||\mathcal{O}_{1}+\mathcal{O}_{2}||\Theta_\alpha^{I}  \rangle \\
+&\sum_{\alpha,\alpha',\nu}c_{1,\alpha'\nu}^*c_{1,\alpha\nu}
(-1)^{J'+J_c+I+1}\hat{I}'\hat{I}
\left\{
\begin{tabular}{ccc}
$J'$ & $I'$ & $L_\nu$ \\
$I$ & $J$ & $1$
\end{tabular}
\right\}\\
\times&\langle \Theta_{\alpha'}^{J'}||\mathcal{O}_{1}+\mathcal{O}_{2}||\Theta_\alpha^{J} \rangle
+\sum_{\alpha,\nu,\nu'}c_{1,\alpha\nu'}^*c_{1,\alpha\nu}\\
\times&(-1)^{J+L_\nu+I'+1}\hat{I}'\hat{I}\left\{
\begin{tabular}{ccc}
$L_{\nu'}$ & $I'$ & $J$ \\
$I$ & $L_\nu$ & $1$
\end{tabular}
\right\}
\langle \Phi_c^{L_{\nu'}}||\mathcal{O}_{\xi}||\Phi_c^{L_\nu} \rangle.
\end{split}
\label{amp}
\end{equation}

The first term of the right hand side in Eq.~\eqref{amp} is the contribution from the valence two nucleons when the core nucleus is inert and the second term corresponds to the higher order correlation due to the core excitation. The third term is the contribution from the core nucleus directly excited by the one-body external field.

\nocite{*}

\bibliography{3bm}

\end{document}